# Propagating Surface Plasmons with Interference Envelope and A Vision for Time Crystals[1,*]

AMIR DJALALIAN-ASSL[1,*]

[1] *51 Golf View Drive, Craigieburn, VIC 3064, Australia*
*\*amir.djalalian@gmail.com*

**Abstract:** The influence of the film thickness and the substrate's refractive index on the surface mode at the superstrate is an important study step surrounding their propagation mechanism. A single sub-wavelength slit perforating a thin metallic film is among the simplest nanostructure capable of launching Surface Plasmon Polaritons on its surrounding surface when excited by an incident field. Here, the impact of the substrate and the film thickness on surface waves is investigated. When the thickness of the film is comparable to its skin depth, SPP waves from the substrate penetrate the film and emerge from the superstrate, creating a superposition of two SPP waves, that leads to a beat interference envelope with well-defined loci which are the function of both the drive frequency and the dielectric constant of the substrate/superstrate. As the film thickness is reduced to the SPP's penetration depth, surface waves from optically denser dielectric/metal interface would dominate, leading to volume plasmons that propagate inside the film at optical frequencies. Interference of periodic volume charge density with the incident field over the film creates charge bundles that are periodic in space and time.

## 1. Introduction

Surface Plasmons (SP), or Surface Plasmon Polaritons (SPP), is a designation given to the quasi-particles representing the collective oscillation of the surface charges at a metal/dielectric interface. The necessary condition for the excitation of the SPPs modes, from the material point of view, is the existence of a 2D electronic gas formed at the interface between a conductor with complex permittivity and a dielectric with positive permittivity. Collective oscillations of surface charges manifest themselves in the form of longitudinal surface waves that propagate along the metals/dielectric interface. Here, the term 'Polaritons' suggests the coupling between the polar excitations (i.e. positive/negative charge bundles) of the surface waves to the electromagnetic fields near the surface that also travel with the SPP[1]. One way of launching SPPs over a metallic surface is the extraordinary optical transmission through a subwavelength hole perforated in a metallic film[2-5]. In a most relevant report (to a certain extent) Wang *et. al.*[6] modelled a free standing optically thin silver film in vacuum, where authors try to explain their findings in terms of long range SPPs, SPP Wave Packets and Quasi Cylindrical Waves (QCW) … etc. Prior to that, Verhagen *et. al.* showed that guided waves in a metal-dielectric-metal waveguide can penetrate the thin metallic cladding hence shortening the wavelength of the SPPs at the silver/air interface [7]. Note that this report is not concerned with the SPP eigenmodes [8-10], but rather it is an investigation on surface wave interference under the forced vibration. Eigenmode analysis identifies modes that are "supported" but not necessarily activated. Given this report being concerned with SPPs launched under the "forced vibration", with a single wavelength and a series of defined film thicknesses, it is inconsequential to distinguish (if that would be possible at all) between any of the SPP modes. In this case, FFT mode analysis is the best approach.

The numerical results reported here are exactly those included in chapter 10 of my thesis [1]. However, in this report, notions such as Lorentz force and periodic transparency are used to explain the mechanism behind the plasmonic time-crystal and more. In this report, I will examine the SPPs launched by a single aperture perforated in a silver film supported on a substrate, firstly glass and later diamond. Schematics for the model under investigation is depicted in Figure 1. Starting from basics I will report on the following effects:

**1**) When the refractive index of the substrate differs from that of the superstrate (i.e. air), SPP waves from the substrate penetrate the film and emerge from the air/silver interface. Consequently, the superposition of the two waves from both sides of the film leads to propagating SPP waves modulated by a well-define non-propagating interference envelope at the silver/air interface. **2**) For a sufficiently thin metallic layer, SPPs formed at the metal/superstrate interfere with those formed at the metal/substrate within the metal leading to a non-propagating periodic electric polarization inside the film. However, although the penetration depths into the silver film are nearly equal from silver/air, silver/glass and silver/diamond interfaces, the surface waves from the optically denser dielectric/metal interface would dominate film. This is purely due to the stronger SPP field localized at the optically denser substrates. **3**) How the periodic electric polarization inside the film would interact with the incident field that would lead to the formation of super-atoms that are periodic in both space and time with a periodicity other than that of the drive. All simulations are performed with incident wavelength of $\lambda_0 = 700$ nm associated with the emission from nitrogen vacancies in diamonds at room temperature [11, 12].

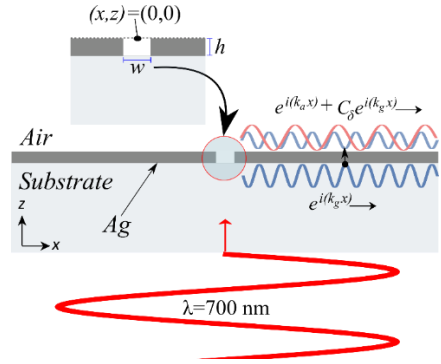

Figure 1: Schematics for the model under investigation

Let us start with the fundamental equations governing the complex wavevectors of SPP stated here for convenience [13]:

$$k_{SPP} = \sqrt{\frac{\varepsilon_m \varepsilon_d}{\varepsilon_m + \varepsilon_d}} k \equiv k'_{SPP} + i\, k''_{SPP} \qquad (1)$$

$$k_m \approx \sqrt{\frac{\varepsilon_m^2}{\varepsilon_m + \varepsilon_d}} k \equiv k'_m + i k''_m \qquad (2)$$

where $k_{SPP}$ is the complex wavevector of the SPPs propagating in the *x*-direction at the metal/dielectric interface with $k_m$ being the SPPs' complex wavevector, propagating in the *z*-direction into the metal respectively. Further details on surface plasmons may be found in the literature [5, 14, 15].

Now, consider a metallic thin film with its surface set parallel to the *x-y* plane. Equation (1) describes the complex wave vector for the SPP waves propagating at the metal/dielectric interface along the *x-y* plane, whereas the wave vector for the SPP waves penetrating the metallic film in the *z*-direction is given by equation (2). In both equations the real part of the wave vector represents propagation constant, whereas the imaginary part defines the decay lengths, $1/k''_{SPP}$ and $1/k''_m$, over which the SPP's amplitude decreases by $1/e$, when propagating over the surface and into the metal respectively. Note that in equation (2), the permittivity, $\varepsilon_d$, corresponds to the dielectric material from which the field penetrates the film.

The short introduction above was aimed to highlight some of the key features of surface plasmon polaritons relevant to this report. What will follow is a theoretical study on SPPs launched by a single subwavelength aperture perforated in a silver thin film. Section 2.1 covers the influence of the film thickness and the refractive index of the supporting substrate on the SPPs and in section 2.2 a plasmonic time crystal is proposed.

## 2. Results and Discussions

### 2.1 The Origin of Modulating Envelope in SPPs over Flat Metallic Films

A 2D Finite Element Method (FEM) model of a 100 nm thick silver film perforated with a 50 nm wide slit was simulated using COMSOL Multiphysics/RF module (EM Wave, Frequency Domain)/Stationary Solver. Model consisted of substrate/silver/air layers with outer boundary terminated with scattering boundary condition (SBC) or perfectly match layers (PML), both of which produced the same results. Structure was illumination from the substrate's SBC being configure with incident electric field operatized the $x$-direction. Modelling time harmonics with FEM is particularly useful in examining the steady-state response of the system under the continues excitation with an incident wave with a single wavelength. The refractive index of the glass substrate supporting the film was initially set to $n_1 = 1.52$ and the refractive index data for silver was taken from Palik [16]. The film was along the $x$ plane and was illuminated with a normally incident TM wave propagating in the $+z$-direction from glass substrate. For convenience the air/silver interface is denoted by $z = z_0$. Figure 2(a) depicts the distribution of the real part of surface charge densities, $\sigma(x,t) = |\sigma(x)| e^{i(k_{SPP}x - \omega_0 t)}$, at an arbitrary time $t_0$, calculated at both the air/silver and glass/silver interfaces from the normal to the surface, i.e. the $z$-component of the electric field. The amplitude, i.e. the envelope, of the surface charge density at the air/silver interface was calculated using $|\sigma(x)| = \sqrt{\sigma(x,t)\sigma(x,t)^*}$ and is depicted in Figure 2(b). Here, surface charge densities, $\sigma = \varepsilon_0 (E_{z1} - E_{z2})$, are calculated using the normal to the surface component of the electric fields $E_{z1}$ and $E_{z2}$, obtained from both sides of the metal/dielectric interface. The corresponding Fast Fourier Transforms (FFT), $f[\sigma(x,t_0)]$ and $f[|\sigma(x)|]$ were also calculated, see Figure 2(c)-(d). In Figure 2(b), the maximum accumulated charge density at the edge of the cavity, $x_1 = 25$ nm, is labelled $C_{max}$. The decay length of the surface charge density, where the value of the $C_{max}$ drop by $1/e$, was found to be ~10 nm from the edge (or 35 nm from the center). At $\lambda_0 = 700$ nm, the decay length of an SPP along the silver/air interface is ~67 μm [13]. Activities near the slit, therefore, may not be considered as SPPs as they are highly localized. The inset of Figure 2(b), depict the $|\sigma(x)|$ and $\sigma(x,t)$ at $t = t_0 + T/6$ and $t_0 + T/4$. Here, $T = 1/f_0$ is the time period associated with the incident EM wave with a drive frequency $f_0 = c/\lambda_0$, and $t_0$ was set to a time when the surface charge density was at its maximum, $C_{max}$, at $x_1$. The separation between the localized surface charges and the appearance of the harmonic wave occurs at $t = t_0 + T/6$ and $x_2 \approx 75$ nm, i.e. 50 nm away from the edge. In fact, the 10 nm decay length, closer to the $1/k''_m \approx 25$ nm obtained from equation (2), indicates that the surface charges in the vicinity of the slit are due to the cavity modes, penetrating the metal and subsequently decaying rapidly. This agrees to previous works [17, 18]. Furthermore, at $t = t_0 + T/4$ the surface charge density at $x_1$ drops to 0 and the peak at $x_3 = 200$ nm resembles that of a harmonic wave. The phase difference of 90° between the oscillations at $x_1$ and $x_3$, resembles that of a forced vibration where the force leads the displacement by 90° under resonance conditions [19]. However, the amplitude of the first peak at $x_3$ is $1.9 \times C_{0a}$, where $C_{0a}$ is the DC component of $|\sigma(x)|$, hence the average amplitude of the propagating SPP waves, see Figure 2(a) and (d). Note FFT of the envelope, $f[|\sigma(x)|]$, in Figure 2(d), identifies the DC components (or the amplitudes of the SPP waves), $C_{0a}$ and $C_{0g}$ at both interfaces.

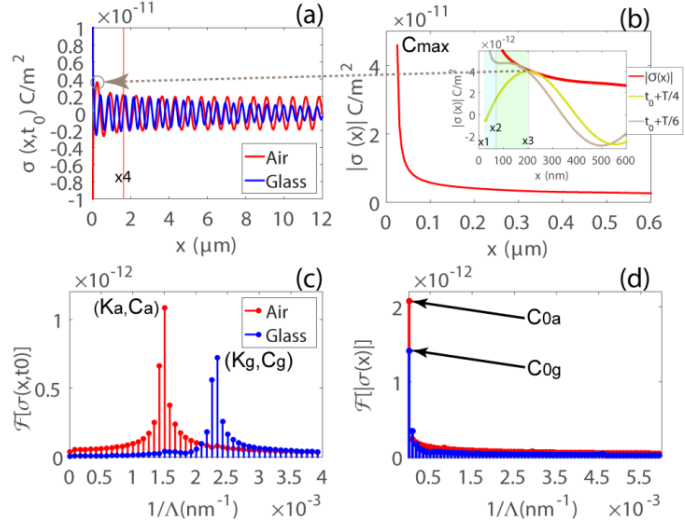

Figure 2: (a) Surface charge density, $\sigma(x, t_0)$, at an arbitrary time $t_0$, calculated at the air/silver and glass/silver interfaces. (b) The envelope, $|\sigma(x)|$, at the air/silver interface. The corresponding FFT of (c) the wave $f[\sigma(x,t_0)]$ and (d) the envelope $f[|\sigma(x)|]$ [1]. $\Lambda$ and $1/\Lambda = K_{SPP}$ are wavelength and wavenumber obtained from FFT respectively.

By examining Figure 2(a) it was determined that the amplitude of the wave drops to $C_{0a}$ at $x_4 \approx w/2 + 2 \times \lambda_{SPP}$, i.e. 2 wavelengths away from the edge of the slit. Although the surface charge density resembles that of a harmonic oscillation in the $x_3 \leq x \leq x_4$ range, its rapid decay and non-conformance to the $1/k''_{SPP}$, suggests a kind of transient state. To evaluate the $\lambda_{SPP}$, FFT transform $f[\sigma(x,t_0)]$ was calculated for both the silver/air and the silver/glass interfaces, Figure 2(c). The weighted average, $K_{SPP} = \sum_{i=1}^{5}(K_{SPPi}C_{SPPi})/C_{SPPi}$, that included the center mode and the four immediate neighboring modes, i.e. two on each side of the maxima, provides a good estimate of SPP wavenumbers. The SPP wavelengths were then calculated using $\lambda_{SPP} = 1/K_{SPP}$, where $K_{SPP} = \text{Re}(k_{SPP})/2\pi$ is the wavenumber obtained from FFT.

For the sake of brevity in notations, let the subscripts "$a$" and "$g$" be denoting the association of physical quantities carried by the SPP waves at the superstrate (air) and substrate (glass and later diamond) respectively. So, in summary, $\lambda_a = 1/K_a = 667$ nm and $\lambda_g = 1/K_g = 427$ nm are in agreement with $\lambda_a = 682$ nm and $\lambda_g = 433$ nm obtained analytically using equation (1). Examining the $|\sigma(x)|$, an additional spatial second harmonic were observed in the envelope at both interfaces. The second harmonics in the envelope seems to be the result of superposition of two time-harmonic waves:

$$\sigma(x,t) \propto Ae^{i(k_a x - \omega_0 t)} \pm Be^{i(k_a x - \omega_0 t + m\pi)} \tag{3}$$

where $m$ must be an even integer and $A \gg B$. However, the origin of the second term in equation (3), $B$, is unknown. The boundary conditions were set to eliminate all reflections, therefore, simulation artefacts cannot account for such periodic perturbations, even more so that such second harmonics do not manifest themselves over the surface of a Perfect Electric Conductor (PEC) that does not support SPPs!

A possible scenario that may lead to oscillations at double the fundamental frequency in $|\sigma(x)|$, is the normal-to-the-surface component of SPPs being modulated by the parallel-to-the-surface component at the interface via a relationship that involved multiplication. SPPs are longitudinal waves manifested as surface charge bundles, where charges in each bundle are held together by SPP's $E_z$ along the $x$-axis. Repelling/attracting Coulomb forces from each bundle to its neighboring charge bundles of equal/opposite signs, is analogous to a chain of masses attached to one another by springs. Given that the oscillation along the chain is being driven by $F \propto \sigma(x,t) E_x(z_0,t)$ from the aperture, it is plausible to attribute the origin of the backward propagating term in equation (3) to $F_x \propto e^{-i2(k_a x - \omega_0 t)}$, where the push/pull by $F_x$ generates the backward propagating waves. Basically, the force modulates the amplitude of the surface charge density wave over $T/2$, during which the SPP has travelled a total distance of $\lambda_g/2$. Having noted that, the exact form of a partial differential equation governing the forced vibration that leads to equation (3) as a solution is yet to be determined.

Regardless, it was envisaged that by reducing the film thickness, it would be possible for surface charge densities from the glass/silver interface manifest themselves at the air/silver interface, leading to a superposition of the two waves:

$$\sigma(x) = \left[ 2 e^{i\left(\frac{(k_a + k_g)x}{2}\right)} \cos\left(\frac{(k_a - k_g)x}{2}\right) \right] \tag{4}$$

see Appendix A. This would modulate the charge densities along the $x$-direction, resulting in a series of minima/maxima with fixed loci that are $1/K_{\text{beat}}$ apart, where $K_{\text{beat}} \equiv |K_a - K_g|$. Hence by controlling the film thickness and the refractive index of the substrate, one could control the modulation strength and frequency of the envelope. Keeping the superstrate and the substrate intact as before, two additional simulations, with $h = \{50, 25\}$ nm, were carried out in order to investigate the influence of the film thickness. Figure 3 depict the numerically calculated $f[\sigma(x,t_0)]$ and $f[|\sigma(x)|]$. Figure 3(a), (c) and (e) depicts $f[\sigma(x,t_0)]$ with $h = \{50, 25\}$ nm when the film is supported on a glass substrate and with $h = 25$ nm on a diamond substrate respectively. In all cases, $K_a$ was found to be at the same position as it was for $h = 100$ nm. For $h = \{50, 25\}$ nm on a glass substrate, $K_g$ was also found to be at the exact location as it was for the 100 nm thick silver film. In the case of the diamond substrate, $\lambda_g = 1/K_g = 230$ nm, was found to be close to the $\lambda_g = 246$ nm calculated using equation (1). In all cases, the appearance of an additional peak at the air/silver interface, positioned at $K_g$ having an amplitude $C_{\delta g} = C_g e^{-\frac{z}{\delta}}$, corresponded to the SPP waves that travel along the substrate/silver interface penetrating the film and emerging at the air/silver interface. Presence of SPPs with wavelength $\lambda_g$ at air/silver interface is significant as it impacts the design criteria for plasmonic metasurfaces. FFT of the corresponding envelopes, $f[|\sigma(x)|]$, in Figure 3(b), (d) and (f), show the anticipated modulating envelope with $K_{\text{beat}} \equiv |K_a - K_g|$. Note that in order to shift the $K_{\text{beat}}$ to overlap with the second harmonics observed in the envelope, the required value for the substrate's refractive index was found to be (see Appendix A) $n_1 = 2.41$ at $\lambda_0 = 700$ nm that corresponds to diamond [20]. With the recent advances in nano-diamond technology, use of diamond substrate is both feasible and practical [21]. Therefore, an additional simulation was carried out with a 25 nm thick silver film supported on a diamond substrate.

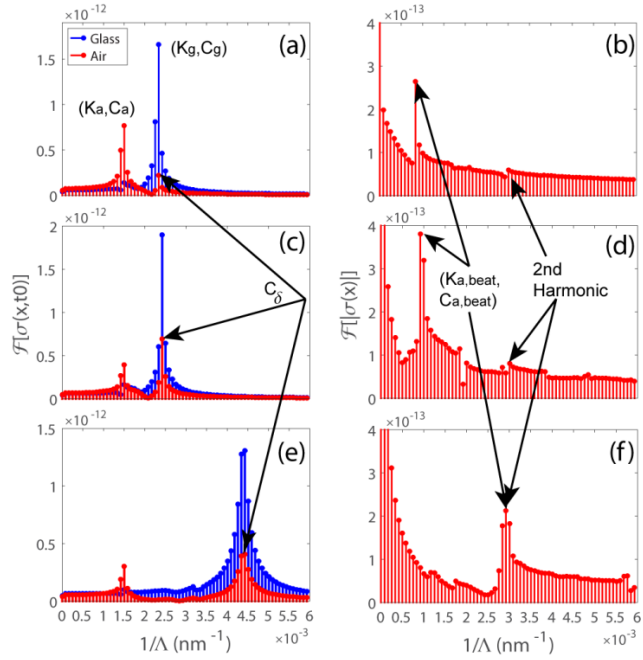

Figure 3: $f[\sigma(x,t_0)]$ and $f[|\sigma(x)|]$ calculated for (a)-(b) $h = 50$ nm on glass substrate, (c)-(d) $h = 25$ nm on glass substrate and (e)-(f) $h = 25$ nm on diamond substrate. Note that subscript 'g' is used to label the substrate in general [1]. $\Lambda$ and $1/\Lambda = K_{SPP}$ are wavelength and wavenumber obtained from FFT respectively.

Figure 4 depicts the modulating envelopes, $|\sigma(x)|$, calculated over the air/silver interface for $h = \{100, 50, 25\}$ nm when the film is supported on a glass substrate and for $h = 25$ nm with a diamond substrate. The aperture was normally illuminated with a Gaussian beam, $15 \times \lambda_0$ in waist, from the substrate. The surface of a perfect electric conductor (PEC) that neither supports SPPs nor allows the penetration of the fields, produced only a smooth line, see Figure 4-(line in black). Values for the PEC line were calculated using $\varepsilon_0 E_z$ to retain the C/m$^2$ unit. The inset in Figure 4 shows the propagating SPPs, $\sigma(x, t)$, that are modulated by the envelope $|\sigma(x)|$, calculated over the air/silver interface for the case $h = 50$ nm when excited with a plane wave from the glass substrate. The presence of the second harmonic and the beat interference in the envelope are marked. A noticeable feature in Figure 4 is the relation between the SPP's decay length, $1/k''_{SPP}$, along the air/silver interface and the strength (or the amplitude) of the interference envelope. Propagating SPP waves along the air/silver interface may be described as the superposition of two waves according to equation (A4) in Appendix A, where each component decay according to their respective decay length $1/k''_a$ and $1/k''_g$. Analytical values for decay lengths were found to be {67, 17, 3.2) μm for the air/silver, glass/silver and diamond/silver interfaces respectively. This explains the decay length of the envelope clearly. For example, in the case of the 25 nm silver film supported on a diamond substrate, the amplitude of the $C_{\delta g}e^{i(k_g x-\omega_0 t)}$ component drops to $1/e$ of its maximum at $x = 3.2$ μm, beyond which the only component that continues to propagate is $C_{0a}e^{i(k_a x-\omega_0 t)}$ due to its longer decay length of ~67 μm. And since the modulating envelope with $K_{beat}$ requires the presence of both components at the air/silver interface, the decay length of the envelope is dictated by the component having the shortest of the two decay lengths, which in this example is 3.2 μm associated with the $C_{\delta g} e^{i(k_g x - \omega_0 t)}$.

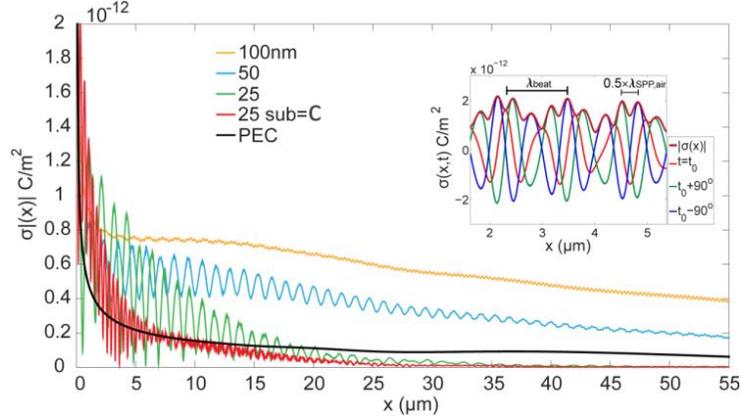

Figure 4: Surface charge densities, $|\sigma(x)|$, over the air/silver surface for $h = \{100, 50, 25\}$ on glass substrate, $h = 25$ nm on diamond substrate and for PEC [1].

Experimental measurements of such effects, however, may not be possible. Although measurements carried out by Verhagen *et. al.* [7] *may* be valid, Wang *et. al.*[6] correctly pointed out that positioning any probe such as an AFM tip, in the vicinity of the slit stablishes standing wave oscillations between the tip and the slit, leading to a series of minima/maxima that convolve with those of the interference envelope. FEM simulations has confirmed this.

The concept of plasmonic microzones[22, 23] are not something new. What sets apart what I have reported here is the formation of the periodic screening/transparency (i.e. the zone plate) by the SPPs alone, and not during the fabrication. For the diamond substrate, diffraction patterns through a 25 nm silver film with single hole when excited with a normally incident beam (Gaussian in *x*) having a waist of $2\times\lambda_0$, is shown in Figure 5(a). Close examination of the light intensity, $|E|^2$, passing through the film, clearly showed the periodic screening/transparency inside the film with spacings $1/K_g$ = 230 nm. I will elaborate on reasons for the SPPs from the substrate dominating the film in the next section. The mechanism, however, that leads to the particular diffraction/scattering pattern in the far-field as seen in Figure 5(a), is mainly due to that of a microzione with interlaced periodic transparent/opaque regions. Basically, the role of the aperture is to partially intercept the incident power in order to launch the SPPs that in turn establish the periodic screening/transparency inside the film, with each opaque-transparent-opaque sequence producing an aperture-like *SPP-induces-transparency*, an effect purely due to the rearrangement of the conduction electrons inside the film. Transparent regions together with the aperture itself form an aperture antenna array. Light transmitted through the array makes constructive interference along the optical axis (i.e. central-lobe) in the far-field as long as the spacing between apertures is less than the transmitted wavelength [24]. The near-field, on the other hand, is dominated by all kind of effects such as envelope-modulated SPPs waves or more. The side-lobes seen in Figure 5(a), are due to the far-field/near-field mixings. When the maximum intensity of the Gaussian beam falls away from the center of the slit (an arbitrary displacement of 680 nm in this case) the intensity of the transmitted beam exhibits a curvature and a tilt towards the displacement, with the transmitted beam being split in two, Figure 5(b). However, the appearance of the curved beam is in fact nothing but strengthening the side-lobe by realigning the maximum incident intensity with the position of the side-lobe, compare the left and right side-lobes in Figure 5(a) to those in Figure 5(b). Such light-matter interaction is not observed in the transmitted beam through a 25 nm silver film on a diamond substrate with no aperture, see Figure 5(c). The dominating mechanism in this case is that of the skin-effect, however, for an explanation governing the transmission through a continuous thin flat silver film in the absent of SPPs see [25].

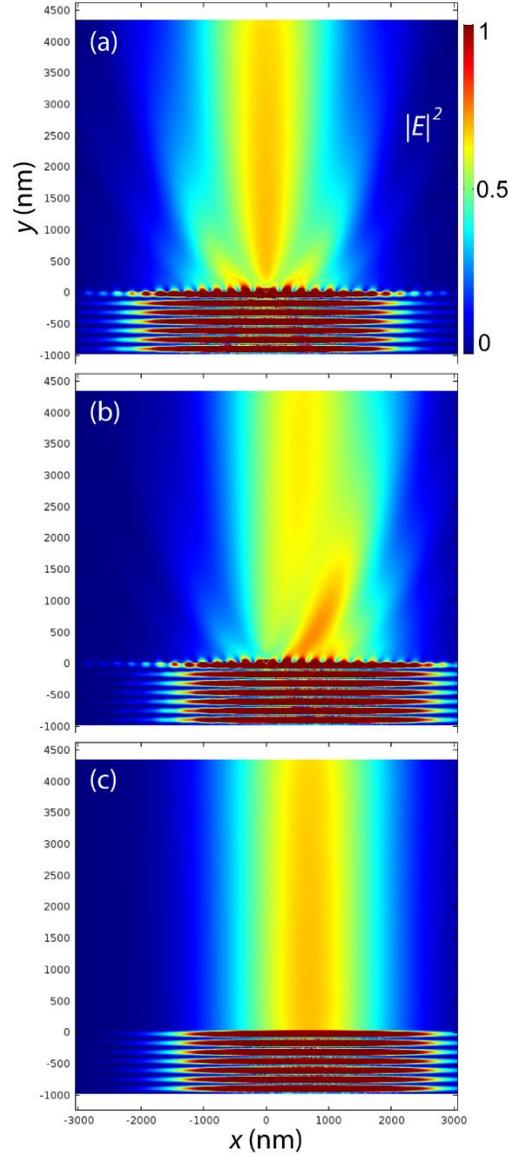

Figure 5: $|E|^2 \times 10 (V/m)^2$ Diffraction patterns of a transmitted Gaussian beam through (a) 25 nm silver film perforated with a slit, supported on a diamond substrate. (b) same as (a) with the maximum intensity of the Gaussian beam displaced to $x = 680$ nm away from the center of the slit. (c) In the absence of the slit [1].

*2.2 Roadmap to Further Study and a Vision for Plasmonic Time Crystal*

With diamond (or glass) substrate, when the film thinness is that of the skin depth, SPPs are no longer confined to the surface of the metal but rather penetrate the film from the substrate and superstrate and interfere with one another inside the film. However, (as it is the case here), due to the aperture dimensions and partially due to the $\varepsilon_m E_{zm} = \varepsilon_d E_{zd}$, normal to the surface component of the electric field at the diamond/silver boundary is much stronger than those at the air/silver at $\lambda_0 = 700$ nm, hence $E_{zg} \gg E_{za}$. Therefore, fields from the substrate dominate the film. Furthermore, the x-component of the SPP was found to be stronger than its z-components at $\lambda_0 = 700$ nm, i.e. $E_{xg} \approx 2E_{zg}$ calculated using $E_{xg} = -i\sqrt{\frac{-\varepsilon_m}{\varepsilon_d}} E_{zg}$, and in agreement with the numerical results. Furthermore a trail of z-component of the electric field $\{\ldots 0 - 0 + 0 - 0 \ldots\}$ carries a trail of x-component $\{\ldots - 0 + 0 - 0 + \ldots\}$ with "+", "0" and "-" denoting $+E_{x,z}$, 0 and $-E_{x,z}$ respectively. Note the 90° phase difference between the x and the z-components. Naturally, the induced periodic polarization, travels inside the film as the SPPs propagate over the surface. The x-component of the polarization, $P_{xg} = \varepsilon_0 \chi_e E_{xg} e^{i(k_g x - \omega_0 t)}$, is of interest in the context of plasmonic time crystals as it signifies periodic accumulation of conduction electrons along the x-axis, hence periodic screening/transparency within the film that resembles that of a Fresnel zone plate.

With a flat metallic film that extends to infinity in the x-direction, it is not possible to apply the Gauss's law to calculate the charges due to SPP fields. Therefore, I have provided an alternative approach to calculate the induced periodic charge density due to propagating $E_{zg}$:

$$\Delta \rho_x(k_g, \omega_0) = -\frac{\omega_0}{c} \varepsilon_0 \varepsilon_m^{"} \left[ \frac{\varepsilon_d + \varepsilon_m}{\sqrt{\varepsilon_m + \varepsilon_d}} \right] E_{zg} e^{i(k_g x - \omega_0 t)} \tag{5}$$

see Appendix B. And in terms of number of electrons being displaced:

$$\Delta N(k_g, \omega_0) = \frac{\Delta \rho_x}{e^-} = -\frac{\omega_0}{ce^-} \varepsilon_0 \varepsilon_m^{"} \left[ \frac{\varepsilon_m + \varepsilon_d}{\sqrt{\varepsilon_m + \varepsilon_d}} \right] E_{zg} e^{i(k_g x - \omega_0 t)} \tag{6}$$

Equation (6) reveals the total number of charges being displaced per SPP field. Given that the fermi energy level for metals is given by $\mathcal{E}_F = \frac{h^2}{8m} \left( \frac{3N}{\pi} \right)^{2/3}$ and the plasma frequency by $\omega_p = \sqrt{\frac{Ne^2}{\varepsilon_0 m_e}}$ [26], one must set $N = N_0 + \Delta N$, where $N_0$ is the number of electrons per unit volume when unperturbed, leading to:

$$\mathcal{E}_F(k_g, \omega_0) = \frac{3^{2/3} h^2}{8m\pi^{2/3}} (N_0 + \Delta N)^{2/3} \tag{7}$$

$$\omega_p(k_g, \omega_0) = \sqrt{\frac{(N_0 + \Delta N)e^2}{\varepsilon_0 m_e}} \tag{8}$$

Consequently, $\Delta \rho_x$, $\Delta N$, $\omega_p$ and $\mathcal{E}_F$ become a function $(k_g, \omega_0)$ due to $E_{zg} e^{i(k_g x - \omega_0 t)}$, which may lead to many interesting effects, such as periodic refractive index, fermi levels, local work functions, density of states, eigen energies inside film, which will be a topic of another report. Nevertheless, in the absence of any incident field over the film, for example when the SPPs are launched by a dipole near the surface [27], the propagation of surface waves, and all physical quantities they carry, is unperturbed. However, in the presence of an incident field normal to the surface, the superposition of the fields inside the film is given by $E = E_{xi} e^{i(k_i z - \omega_0 t)} + E_{xg} e^{i(k_g x - \omega_0 t)}$ which create disturbance on the periodic charges densities. It is intuitive that loci polarized by $+E_{xg}$ be transparent to $+E_{xi}$, and vice versa. Now, consider an arbitrary time $t = t_0$, when the maximum of the incident electric field falls over the film. This is depicted by the following notation:

$$\frac{E_{xg}(x,t_0): +,0,-,0,+}{E_{xi}(t_0): +,+,+,+,+} \Rightarrow +,+,0,+,+ \tag{9}$$

This scenario is shown in Figure 6 with the periodic arrangement of "0"s when the maximum of the field falls over the film, supported on a glass and diamond substrates respectively. For a given incident wavelength (or a given drive frequency) and the choice of metal, the spacing between "0"s are the function of substrate's refractive index. Also note the strong periodic field under the film, inside the glass and diamond substrates! At $t = t_0+T/2$, hence 180° phase, both $E_{xg}$ and $E_i$ change sings. This will lead to:

$$\frac{E_{xg}(x,t_0+T/2): -,0,+,0,-}{E_{xi}(t_0+T/2): -,-,-,-,-} \Rightarrow -,-,0,-,- \tag{10}$$

with "0" remained intact in space as it was in the case of equation (9). This scenario is also confirmed by numerical results.

*It is intuitive to think of the periodic "0"s as loci where conduction electrons are trapped. If this hypothesis is validated by experiment, it would open doors to study new phenomena. Each "0" may be viewed as a super-atom with oversaturated electronic orbitals, elevated fermi level, lowered work function and many more effects when considered in periodic settings which I have highlighted in the conclusion. The perpetual creation and annihilation of such plasmonic super-atoms being periodic in space and time with frequencies other than that of the drive, never attaining thermal equilibrium, qualifies them as a plasmonic time-crystal as I shortly elaborate.*

Back to causality, at the first glance it seems it is a simple matter of superposition of two orthogonally propagating EM waves with the *x*-component of the two fields summed up inside the film. However, a close look at the numerical results revealed that when $E_{xi}$ drops to 0, (e.g. at $t = t_0+T/4$), polarization inside the film experiences the effect of $E_{xi}(t_0)$. This 90° phase difference between applied field and the reaction is attributed to the charge bundles experiencing the Lorentz force $F_{xg} = J_{zg}(B_{yg} + B_{yi})$. Furthermore, since $B_{yi}$ is 0 at $t = t_0+T/4$, the restoration of periodic potential is resumed at $t = t_0+T/4$ but completed at $t = t_0+T/2$ when $-E_{xi}$ falls over the film. The whole creation/anhelation of $P_x = \varepsilon_0 \chi_e E_x$ is a sinusoidal process in time. Numerical results also revealed that $E_{zg}$ not being affected by the *incident* field. Therefore, the restoration process is attributed to the $E_{zg}$. Since this creation/anhelation is periodic both in time and space with periodicities $T/2$ and $\lambda_g$, hence oscillating at frequency other than that of the drive, (although not an expert in the topic), I believe it qualifies as a time crystal [28, 29]. As for the breaking time symmetry, I need access to certain resources only academics enjoy, but as an alumnus, this is not possible at this stage. However, I must remind the readers that the *x*-component of the SPP's electric field (as I understand) must always lag its *z*-components by 90° at *resonance*, a condition satisfied for SPPs launched by an aperture at any frequency.

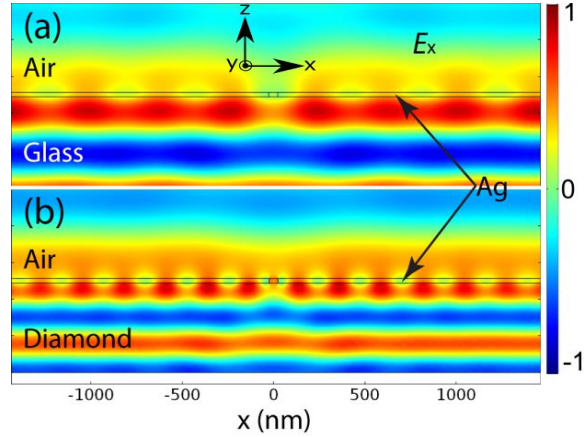

Figure 6: Snapshot of electric field $E_x$ passing through a periodic charge screen (with periodicity $1/K_g$) formed inside the 25 nm thick silver film for (a) glass and (b) diamond substrates. Note that $E_x$ was calculated at an arbitrary time with the maximum of its amplitude falling over the silver film [1].

The term *crystal* also implies that one should be able to define both the Hamiltonian to determine the eigen-energies and the Schrödinger equation to explain the De Broglie's matter waves [30, 31] for that system. The potentials $V$, experienced by electrons in a time crystal and consequently, wave functions $\psi$ and eigen-energies $\xi$, must naturally be time-dependent:

$$\xi_j(t)\psi_j(\mathbf{R}_j,t) = \hat{H}(t)\psi_j(\mathbf{R}_j,t) \tag{11}$$

$$i\hbar\frac{\partial \psi_j(\mathbf{R}_j,t)}{\partial t} = \hat{H}(t)\psi_j(\mathbf{R}_j,t) \tag{12}$$

Considering the original Hamiltonian, $\hat{H}$, for the many-electrons [32]:

$$\hat{H} = \sum_j \left[ -\frac{\hbar^2}{2m}\nabla_j^2 + V(\mathbf{R}_j) \right] + \frac{1}{2}\sum_{\substack{j,k \\ j\neq k}} \frac{e^2}{4\pi\varepsilon |\mathbf{R}_j - \mathbf{R}_k|} \tag{13}$$

Time variations of kinetic energy term in equation (13) is taken care of by the time-dependent wave function $\psi_j(\mathbf{R}_j,t)$, however, one must introduce the notion of time into equation (13) and rewrite it as:

$$\hat{H}(t) = \sum_j \left[ -\frac{\hbar^2}{2m}\nabla_j^2 + V(\mathbf{R}_j(t)) \right] + \frac{1}{2}\sum_{\substack{j,k \\ j\neq k}} \frac{e^2}{4\pi\varepsilon |\mathbf{R}_j(t) - \mathbf{R}_k(t)|} \tag{14}$$

where $V(\mathbf{R}_j(t)) = V_{ext}(\mathbf{R}_j(t)) + V_t(\mathbf{R}_j(t),t)$, $V_{ext}$ is the potential due to the positive ions, $V_t \propto e^{i(k_g x - \omega_0 t)}$ is the time-dependent potential due to the creation/anhelation of charge bundles and $\mathbf{R}_j$ is the position vector of the $j^{th}$ electron. Note that by making $\mathbf{R}_{j,k}$ time-dependent, $j \neq k$ is taken care of, however, $V_t$ has both time and spatial dependence (other than that of positive ions). Time-dependent Hamiltonian in Equation (14) implies that the computation must trace the position of each electron, $\mathbf{R}_j$ with respect to $t$ and the changes in potential with respect to $t$ and $\mathbf{R}_j$. The Hartree approximations [32] are also based on time-independent electron-electron interactions, so it must be remedied accordingly for time crystals. An interesting article by

Linde [33] may prove to be useful to investigate possible changes to the effective mass and conductivity in an applied field but that also needs to be modified. With the advent of High Performance Computing (HPC) ab initio modelling and simulations of matters where the constituting components are atoms and electrons are becoming more accessible. An article by Borysov *et. al.* [34] provides a background on the existing infrastructure for numerically modelling and investigating structures at atomic levels using the Density Functional Theory (DFT) calculations, which may prove to be a platform of choice to study time crystals.

## 3. Conclusions

In conclusion, it was shown that for a sufficiently thin silver film sandwiched between two different dielectrics, the mixing of the two SPPs (formed at the substrate and the superstrate) produce an interference envelope that modulates the propagating SPPs. For film thicknesses equivalent to the SPP's penetration depth, surface waves from optically denser dielectric/metal interface would dominate, leading to volume plasmons that propagate inside the film at optical frequencies. Interference of such volume charges with the incident field over the span of the film creates charge bundles that are periodic in space and time. Although many questions remained unanswered in this report, the future work will focus on them. I would hypothesize that the presence of charge bundles inside the film may imply changes to the electronic density of states, electron-electron collision (hence the mean free path), electron-lattice interaction (hence the electron's effective mass) and consequently conductivity, due to the presence of an additional periodic potential that may compete or superpose with that of the positive ions. It is intuitive to think of the periodic "0"s as loci where electrons trapped. If this hypothesis is validated by experiment, it would open doors to study new phenomena. Each "0" may be viewed as super-atom with oversaturated electronic orbitals, elevated fermi level, lowered work function and many more effects when considered in periodic settings which are analogous to that of a superlattice in semiconductors[32].acknowledgments and disclosures

*Conflicts of Interest:*

The authors declare that there are no conflicts of interest related to this article. Apart from the named author, no other person or entity had any role in the inception or the design of the study; in the collection, analysis, or interpretation of data; in the writing of the manuscript, and in the decision to publish the results.

## Appendix A – Superposition

To explain the overlap between the beat modulation and the second harmonics in the envelope when $n_1 = 2.41$ at $\lambda_0 = 700$ nm, consider the superposition of two waves having equal amplitudes propagating along the *x*-axis i.e. $\Psi(x,t) = e^{i(k_1 x - \omega_1 t)} + e^{i(k_2 x - \omega_2 t)}$, which can be written as:

$$\Psi(x,t) = 2e^{i\left(\frac{(k_1+k_2)x}{2} - \frac{(\omega_1+\omega_2)t}{2}\right)} \cos\left(\frac{(k_1-k_2)x}{2}\right)\cos\left(\frac{(\omega_1-\omega_2)t}{2}\right) \tag{A1}$$

Here it is assumed both waves start in phase. This form of the equation is of interest since it separates the terms related to the coherent length, $4\pi/(k_1+k_2)$, and the coherent time, $4\pi/(\omega_1+\omega_2)$ of the superposed propagating wave. Furthermore, the last two cosine terms indicate that the combined propagating wave is modulated by two envelopes having nodes (or anti-nodes) separated by $\left|\cos\left(\frac{(k_1-k_2)x}{2}\right)\right|$ in space and $\left|\cos\left(\frac{(\omega_1-\omega_2)t}{2}\right)\right|$ in time. In other words the beat frequencies in space and time are $|k_1 - k_2|$ and $|\omega_1 - \omega_2|$ respectively, therefore the coherent lengths (for a lack of a better word) of the envelopes in space and time can be calculated as $2\pi/(k_1 - k_2)$ and $2\pi/(\omega_1 - \omega_2)$ respectively. Necessary conditions to eliminate undesirable jitters in space and time envelopes are:

$$[2\pi/(k_1 - k_2)] = [4\pi/(k_1 + k_2)] \quad (A2)$$

AND

$$[2\pi/(\omega_1 - \omega_2)] = [4\pi/(\omega_1 + \omega_2)] \quad (A3)$$

In the case of two superposed SPP waves at the air/silver interface, the superposition may be simplified to:

$$\sigma(x,t) = e^{-i\omega_0 t}\left[C_a e^{i(k_a x)} + C_{\delta g} e^{i(k_g x)}\right] \quad (A4)$$

Furthermore, in the case of 25 nm silver film $C_{\delta g} \approx C_a$, see Figure 3(c) and (e). Under such conditions, equation (A4) may be written for the spatial terms as:

$$\sigma(x) = \left[2e^{i\left(\frac{(k_a + k_g)x}{2}\right)} \cos\left(\frac{(k_a - k_g)x}{2}\right)\right] \quad (A5)$$

Equation(A5), which is the special case of equation(A1), shows the coherent length of the combined propagating SPP waves to be $4\pi/(k_a + k_g)$ with the beat modulation occurring according to $2\pi/|k_a - k_g|$. A necessary condition to overlap the coherent length and the beat modulation with the second harmonics in the envelope is then:

$$1/2k_a = 2\pi/|(k_a - k_g)| = 4\pi/(k_a + k_g) \quad (A6)$$

In this report, numerical values for wavenumbers obtained from FFT showed $1/2K_a \approx 1/|K_a - K_g| \approx 2/|K_a + K_g|$ at $\lambda_0 = 700$ nm when $n_1 = 2.41$. Clearly in the case of a glass substrate with $n_1 = 1.52$, $1/2K_a \neq 1/|K_a - K_g| \neq 2/|K_a + K_g|$ at $\lambda_0 = 700$ nm.

The appendix is an optional section that can contain details and data supplemental to the main text. For example, explanations of experimental details that would disrupt the flow of the main text, but nonetheless remain crucial to understanding and reproducing the research shown; figures of replicates for experiments of which representative data is shown in the main text can be added here if brief, or as Supplementary data. Mathematical proofs of results not central to the paper can be added as an appendix.

### Appendix B - Lorentz Force and Induce Charges Inside the Film

Let the magnetic flux density and the electric field carried by SPPs be denoted by:

$$\mathbf{B}_m = (0, B_y, 0)e^{i(k_x x + k_z z - \omega_0 t)} \quad (B1)$$

$$\mathbf{E}_m = (E_x, 0, E_z)e^{i(k_x x + k_z z - \omega_0 t)} \quad (B2)$$

For simplicity, ignoring the exponent terms, we are interested in $F_x = J_z B_y$ where $J_z = \sigma_e E_z$.
Given that $\mathbf{B}_m = \dfrac{\nabla \times \mathbf{E}_m}{j\omega_0}$

and

$$\nabla \times \mathbf{E}_m = \begin{pmatrix} \partial_x \\ \partial_y \\ \partial_z \end{pmatrix} \times \begin{pmatrix} E_x \\ 0 \\ E_z \end{pmatrix} = \begin{pmatrix} \partial_y E_z \\ -\partial_x E_z + \partial_z E_x \\ \partial_y E_x \end{pmatrix} \quad (B3)$$

In 2D:

$$\nabla \times \mathbf{E}_m = \begin{pmatrix} 0 \\ -ik_x E_z + ik_z E_x \\ 0 \end{pmatrix} \quad (B4)$$

One can write $B_y = \dfrac{k_x E_z - k_z E_x}{\omega_0}$. The Lorentz force distribution along the *x*-axis is then $F_x = J_z B_y = \sigma_m E_z \dfrac{k_x E_z - k_z E_x}{\omega_0}$. This can be further reduced by $E_x = -i\sqrt{\dfrac{-\varepsilon_m}{\varepsilon_d}} E_z$ to:

$$F_x = \dfrac{\sigma_m}{\omega_0} \left[ k_x + k_z \left( i\sqrt{\dfrac{-\varepsilon_m}{\varepsilon_d}} \right) \right] E_z^2 \quad (B5)$$

Using equations (1)-(2) and replacing $\sigma_m = \omega_0 \varepsilon_0 \varepsilon''_m$ [35], the Lorentz force becomes:

$$F_x = \dfrac{\omega_0 \varepsilon_0 \varepsilon''_m}{c} \left[ \dfrac{\sqrt{\varepsilon_m \varepsilon_d} + i\varepsilon_m \sqrt{\dfrac{-\varepsilon_m}{\varepsilon_d}}}{\sqrt{\varepsilon_m + \varepsilon_d}} \right] E_z^2 \quad (B6)$$

The volume charge profile along the *x*-direction due to only the Lorentz force may be calculated as $\Delta \rho_x = F_x / E_x$ where $E_x = -i\sqrt{\dfrac{-\varepsilon_m}{\varepsilon_d}} E_z$, therefore:

$$\Delta \rho_x = -\dfrac{\omega_0}{c} \varepsilon_0 \varepsilon''_m \left[ \dfrac{\varepsilon_d + \varepsilon_m}{\sqrt{\varepsilon_m + \varepsilon_d}} \right] E_z \quad (B7)$$

And in terms of number of free electrons:

$$\Delta N = \dfrac{\Delta \rho_x}{e^-} = -\dfrac{\omega_0}{ce^-} \varepsilon_0 \varepsilon''_m \left[ \dfrac{\varepsilon_d + \varepsilon_m}{\sqrt{\varepsilon_m + \varepsilon_d}} \right] E_z \quad (B8)$$